\documentstyle[prl,aps,twocolumn,epsf,epsfig]{revtex}

\begin{document} 

\title{Equilibration Time Scales  in  Homogeneous Bose-Einstein Condensate 
Dynamics}

\author{ 
Daniel G. Barci$^{1}\thanks{E-mail address: barci@uerj.br}$, 
Eduardo S. Fraga$^{2}$\thanks{E-mail address: fraga@th.u-psud.fr},
Marcelo Gleiser$^{3}$\thanks{E-mail address: gleiser@dartmouth.edu} and 
Rudnei O. Ramos$^{1}$\thanks{E-mail address: rudnei@dft.if.uerj.br}
}

\address{
{\it $^{1}$Departamento de F\'{\i}sica 
Te\'orica,}
{\it Universidade do Estado do Rio de Janeiro,}
{\it 20550-013 Rio de Janeiro, RJ, Brazil }\\
{\it $^{2}$ Laboratoire 
de Physique Th\'eorique, Universit\'e Paris-Sud XI, 
B\^atiment 210, F-91405 Orsay, France}\\
{\it $^{3}$Department of Physics and Astronomy, Dartmouth College,}
{\it  Hanover, NH 03755-3528}
}

\maketitle

\begin{abstract} 

We study the nonequilibrium growth of a weakly interacting
homogeneous Bose gas after a quench from a high-temperature 
state to a temperature below the Bose-Einstein critical 
condensation temperature.
We quantitatively characterize the departure from thermal
equilibrium and observe the presence of {\em two equilibration time scales}. 
The equilibration 
times are shown to be inversely proportional to the density.
\noindent 
PACS number(s): 05.70.Ln, 05.70.Fh, 64.60.Ak
\end{abstract} 

\vspace{0.25cm}

\centerline{\large \bf In Press Physica A (2002)}
\centerline{\bf cond-mat/0110447}

\vspace{0.25cm}

The nonequilibrium evolution of a Bose-Einstein condensate is a
fascinating subject not only on its own but also because it allows
for experimentally probing general ideas concerning the dynamics of
phase transitions. Since the first experiments with dilute atomic gases
led to the accomplishment of Bose-Einstein condensation
\cite{experiments}, a great theoretical and experimental effort
has been put in the investigation of different aspects of equilibrium
and dynamical properties of the Bose-Einstein condensate \cite{reviews}.
To date, however, a comprehensive description and complete understanding
of the nonequilibrium behavior of the condensate is still lacking. In
this context, only a few experiments involving the study of the {\it growth}
of the condensate in trapped dilute atomic gases \cite{growth1,growth2} have been
performed. {}From a theoretical standpoint, different growth
equations, based on the Boltzmann equation applied to inhomogeneous
trapped Bose gases, were proposed \cite{stoof,gardiner}. 

Recently, some of us proposed a method based on the closed-time path
nonequilibrium field theoretical technique to study the growth of the
{\em homogeneous} Bose condensate \cite{BFR}. We pointed
out the fundamental role played by the non-condensate fluctuations in
the dynamics of the condensation growth, and were able to describe the
condensate as well as the non-condensate components of the gas in a
unified microscopic formalism without introducing any phenomenological
parameter. However, from a conceptual point of view, the growth of the
condensate is not the only observable that characterizes the
out-of-equilibrium dynamics of a bosonic system. In fact, to fully
describe the out-of-equilibrium dynamics it would be necessary to
compute higher-order correlation functions at 
two different times. 

Hence, an important issue that deserves a more detailed study
concerns the nonequilibrium process following a rapid quench of the
vapor of dilute atomic gases, from a state well above the condensation
temperature to a temperature below critical. Experimentally,
this can be accomplished by the combined effect of a magnetic trap
followed by a rapid RF sweep, as was done in the MIT experiment
described in Ref. \cite{growth1,growth2}. The final equilibrium state of the system
depends strongly on the nonequilibrium evolution after
the quench process. Therefore, it is imperative to have a formalism that
is able to provide not only a description of the formation of the
condensate but also to quantify the nonequilibrium behavior
of the whole system. In another context, Gagne and Gleiser \cite{GG} 
have proposed
a simple way to reliably measure the departure from thermal equilibrium
of a symmetry-broken system which undergoes a temperature quench. The
method makes use of the absolute value of the rate of change of the
momentum-integrated structure function, $\Delta S_{\rm tot}(t)$, which
we will properly define for our case (see Eq.\ (\ref{structure2})
below). The main idea is that for $t<0$ ({\it i.e., } before the
quench) the system is in thermal equilibrium and $\Delta S_{\rm
tot}(t)\approx 0$, since in this regime each single field mode does not
change in time significantly. Similarly, for large times the system
returns to equilibrium (with a different temperature) and this
quantity again approaches zero. Therefore, any departure from zero of
$\Delta S_{\rm tot}(t)$ is a measure of the out of equilibrium regime.
In other words, $\Delta S_{\rm tot}(t)$ behaves as an ``order parameter''
that is sensitive to the out-of-equilibrium evolution of the system. In
Ref. \cite{GG} this quantity was evaluated for a simple quenched
relativistic model and exhibited a peaked structure. 
It was shown that the height of the peak measures the
departure of the system from thermal equilibrium whereas the location of
the peak is a precise measure of the equilibration time scale, 
$\tau_{eq}$.

In this work, we apply the formalism developed in Ref. \cite{BFR} to
evaluate $\Delta S_{\rm tot}(t)$ in order to study the nonequilibrium
behavior following a quench in a weakly-interacting homogeneous dilute
Bose gas. The system is assumed to be taken from a high temperature
equilibrium state, well above the critical temperature of Bose-Einstein
condensation, $T_i \gg T_c$, to an unstable state in a thermal bath at a
temperature $T<T_c$. Although this may be a strong idealization of the
real setup of recent experiments involving the evolution of dilute
atomic gases subject to a confining potential, we nevertheless expect
that this analysis will provide important qualitative insights to the
actual nonuniform systems. In fact, this approach should
generically be applicable to the dynamics of trapped atomic gases in the
central region of wide traps. As we will show below, 
our results reproduce all qualitative aspects of the condensate
growth. 

Our main result is the determination of a temperature range in which
there are clearly {\em two distinct equilibration time scales}. One
of the equilibration times is essentially determined by thermal
fluctuations dominated by pair correlations, whereas the other is governed
by the condensate growth. We also obtain numerically a scaling 
behavior of these quantities with the density.
  
{}Following Ref. \cite{BFR} we consider a field theory effective model for a 
nonrelativistic complex Bose field, $\phi({\bf x},t)$, with mass
$m$, and a hard core interaction potential, 
$V({\bf x}-{\bf x}')=g \delta({\bf x}-{\bf x}')$,
where the coupling constant, $g$, is related to the $s$-wave scattering
length, $a$, as $g= 4 \pi \hbar^2 a/m$. 
We use the standard decomposition \cite{beliaev} 
of the fields ($\phi,\phi^*$) into
a condensate (uniform) part ($\varphi_0,\varphi_0^*$)
and a fluctuation (nonuniform)
part ($\varphi,\varphi^*$) that describes the atoms (excitations) outside the
condensate, as $\phi({\bf x},t)= \varphi_0 (t) +
\varphi ({\bf x},t)$ and  $\phi^* ({\bf x},t)= \varphi_0^* (t) +
\varphi^* ({\bf x},t)$, where we have assumed a homogeneous condensate.
Particle conservation enforces the constraint on the total
density, 

\begin{equation}
n= |\varphi_0(t)|^2 + \langle \varphi^* \varphi \rangle\;,
\end{equation}
carried throughout the dynamics. Treating the interaction terms in the
self-consistent Hartree-Fock (HF), which is also called
Hartree-Fock-Bogoliubov approximation, we can deduce the 
following equation of motion for the field fluctuations: 

\begin{equation}
i \frac{\partial \varphi}{\partial t} \!=\!
\left(-\frac{\nabla^2}{2 m} - \mu\right) \varphi + 2 g n \varphi +
g (\varphi_0^2(t) + \langle \varphi \varphi \rangle) \varphi^*,
\label{eom}
\end{equation}
\noindent
and an analogous expression for $\varphi^*$. 
The quantities $\langle \varphi \varphi \rangle$ and 
$\langle \varphi^* \varphi^* \rangle$ are usually known as the
anomalous density terms, while $\langle \varphi^* \varphi \rangle$
is the so-called non-condensate density. 

In the mean field (or self-consistent HF) approach used here,
we restrict ourselves to the dynamics in the collisionless 
regime (see e.g. Ref. \cite{giorgini}), valid when the frequency
of the excitations, $\omega_{\rm exc}$, is much larger than the typical
collision (or relaxation) time, $t_{\rm coll}$: 
$\omega_{\rm exc} t_{\rm coll} \gg 1$. In the opposite regime,
$\omega_{\rm exc} t_{\rm coll} \ll 1$, called hydrodynamical regime,
the dynamics happens in the collision-dominated regime in which we
can no longer neglect three-field contributions like 
$\langle \varphi^* \varphi \varphi^* \rangle$ that would enter in the
equation of motion for the condensate and fluctuations. These terms
describe collisional effects that become important for temperatures
close to the critical temperature of transition, and must be included 
in order to reproduce the  results 
of the experiments of condensation dynamics in a trap.
They can be introduced, for example,  in the context
of quantum Boltzmann equations, and lead to a consistent
description of the evaporative cooling and subsequent formation and
equilibration of the condensate, leading to a consistent description
of the experimental data available, as shown in Refs. 
\cite{stoof,gardiner}. As we will discuss below, the parameters
we use fall deeply into the collisionless regime, warranting the
validity of the mean-field approach. 

In Eq. (\ref{eom}),
the chemical potential in the HF approximation is  given by
\cite{griffin} $\mu_{\rm HF}\! =\! g(n + \langle \varphi^* \varphi \rangle 
+\langle \varphi \varphi \rangle )$. It however does not satisfy  
the Hugenholtz-Pines (HP) 
theorem for the chemical potential, $\mu_{\rm HP} \!=\! 
g(n + \langle \varphi^* \varphi \rangle 
-\langle \varphi \varphi \rangle )$, resulting then in the existence
of a gap in the
spectrum of elementary excitations at equilibrium \cite{hohen}, 
in opposition to the
known results for Bose condensates. Even though we treat the nonequilibrium 
evolution of the condensate where the Hugenholtz-Pines theorem,
which is a  purely static relation valid in equilibrium, is not relevant
\cite{stoof2}, the use of the HF approximation for the description of the
noncondensate fluctuations violates the conservation of particle number, 
which in equilibrium results in the violation of the HP theorem.
This is a symptom of the fact
that this approximation does not respect the constraints on the
dynamics imposed by the spontaneous breaking of a $U(1)$ symmetry.

On the other hand, in the Hartree-Fock-Bogoliubov-Popov (HFBP) approximation, 
where the anomalous density is neglected in Eq. (\ref{eom}), the theory
is gapless \cite{hohen}. This is the approximation used in Ref. \cite{BFR},
where the chemical potential was shown to be $\mu = 2 g n $.
One way to go beyond the HFBP approximation and at the same time keep
the theory gapless, as proposed in 
Ref. \cite{burnett}, is to treat the (contact) coupling constant $g$
in the HFBP approximation as an effective  coupling constant $g_{\rm eff}$ that 
can be defined as

\begin{equation}
g_{\rm eff}= g (1+\langle \varphi \varphi \rangle/|\varphi_0|^2)\;.
\label{geff}
\end{equation}
As shown by the authors in Ref. \cite{burnett}, this is a consistent way to 
include the effects of pair correlations,
represented by the anomalous density, and  accounts   
for the low momentum limit corrections of the many body
T-matrix keeping the theory gapless.

Eq. (\ref{eom}) in the improved HFBP approximation together with the 
constraint on the total density
completely determines the time evolution of the condensate, the
non-condensate density $\langle\varphi^*\varphi\rangle$, as well as the
anomalous density $\langle\varphi\varphi\rangle$ in $g_{\rm eff}$.
In order to solve these equations numerically, it is convenient to rewrite 
them in terms of 
their real field components, $\varphi = \varphi_1 + 
i \varphi_2$ and its complex conjugate.
The two-point function $\langle \varphi^* \varphi \rangle$ and 
the anomalous densities can be expressed in a convenient way in terms of the
Green's functions $\langle \varphi_i \varphi_j \rangle$ ($i,j=1,2$) 
constructed from the homogeneous solutions
to the operator of quadratic fluctuations. 
Using the usual Schwinger-Keldysh formalism 
(see, {\it e.g.}, \cite{lebellac,stoof2}) we find \cite{BFR} 

\begin{eqnarray}
\langle \varphi_j (t) \varphi_j (t) \rangle &=& \int \frac{d^3 k}{(2 \pi)^3}
\frac{i}{1-\exp \left(-\beta \varepsilon_{\bf k}
\right)}     
\left[ \chi_j ({\bf k},t)\chi_j^* ({\bf k},t')
\right. \nonumber \\
&+& \left.
\exp\left(-\beta \varepsilon_{\bf k}\right)
\chi_j^* ({\bf k},t)\chi_j ({\bf k},t')\right]\; ,
\label{Gmodes}
\end{eqnarray}
where $\varepsilon_{\bf k}={\bf k}^2/(2 m)$, and $\beta^{-1}$ is the final
temperature the system will reach at long times. The equations for
the fluctuation
modes, $\chi_1$ and $\chi_2$, that are the homogeneous solutions to the
operator of quadratic fluctuations, were derived before \cite{BFR}
in the HFBP approximation. These mode equations, in the improved HFBP 
approximation discussed above, become

\begin{eqnarray}
&&\frac{d \chi_2 }{d t} + \left[\frac{{\bf k}^2}{2 m} +  
g_{\rm eff}  |\varphi_0|^2 \right] \chi_1 = 0 \quad , \nonumber \\
& & \frac{d \chi_1}{d t}  -  \left[\frac{{\bf k}^2}{2 m} - 
g_{\rm eff} |\varphi_0|^2 \right] \chi_2 = 0 \quad ,
\label{modes}
\end{eqnarray}
where $g_{\rm eff}$ is given by Eq. (\ref{geff}).
The boundary conditions for the solution of Eq. (\ref{modes}) are such
that for $t<0$, $|\varphi_0 (t)|^2=0$ and $\chi_1 ({\bf k},t)$ and
$\chi_2 ({\bf k},t)$ are given by $\chi_1 ({\bf k},t) =\exp
(i\varepsilon_{\bf k}t)/\sqrt{2N}$ and $ \chi_2 ({\bf k},t)= i\exp
(i\varepsilon_{\bf k}t)/\sqrt{2N}$ . The normalization factor $N$ is easily
found to be \cite{BFR}
$N=2(\beta_0/\beta)^{3/2}$ where $\beta_0^{-1}$ is the critical temperature
for condensation of an ideal Bose gas.

The density constraint, when expressed as a solution for the
condensate density $n_0 (t) = | \varphi_0 (t) |^2$, becomes
\cite{BFR}

\begin{eqnarray}
|\varphi_0(t)|^2 &=& \frac{1}{2\pi^2} \left(\frac{\beta}{\beta_0}\right)^{3/2}
\int_{k < k_c}
dk~k^2~ \nonumber \\
&\times& \left[ 1-
N \left(|\chi_1({\bf k},t)|^2 + |\chi_2({\bf k},t)|^2\right)\right]
n_{{\bf k}}(\beta) \quad ,
\label{vinculo}
\end{eqnarray}
where $n_{{\bf k}}(\beta)=(e^{\beta\varepsilon_{{\bf k}}}-1)^{-1}$. 
The momentum integral
in Eq. (\ref{vinculo}) is bound to values of $k < k_c$, 
corresponding to the 
unstable modes that drive the system from the initial excited state to the 
condensate state. $k_c$ can be determined from the equation 
for the modes, and is given by the
value of $k$ which changes the sign in the product of the frequencies in 
Eq. (\ref{modes}). 

In {}Fig. 1 we display the result for the evolution of
the condensate density, as given by the solution for the coupled 
integro-differential system of equations (\ref{modes}) and (\ref{vinculo}), 
for three values of the final temperature.
This result improves over the previous one, presented in Ref. \cite{BFR}, 
which was obtained in the HFBP approximation 
by neglecting the contribution from the anomalous densities. 
{}Although the HFBP approximation describes well the condensate for
temperatures well below the critical temperature, it is well known that it
breaks down for temperatures of the thermal bath near the critical one, in
which case the anomalous densities become of the same magnitude 
as the out of the condensate excitation densities. 
In particular, we show in 
{}Fig. 2 that even for moderate temperatures the anomalous densities can
grow to appreciable values during the out-of-equilibrium evolution of the 
system.

{}From the results of {}Fig. 1 we have verified that,
at large $t$, the temperature dependence of the condensate
(in equilibrium) is the expected one, matching the known results for
the drop of the condensate density in equilibrium for increasing 
temperatures in an interacting gas \cite{reviews}. 
However, for initial
times, this tendency changes due to pair correlation contributions
represented by the anomalous density, $\langle \varphi\varphi \rangle$,
in Eq. (\ref{modes}). This fact suggests
the existence of two different scales for the equilibration time: one
related to the condensate growth and the other to the anomalous density. 

In order to investigate this behavior more precisely,  
let us consider the structure function 
defined as the {}Fourier transform of the two-point Green 
function, $S_k(t)=\langle\varphi_j(t)\varphi_j(t)\rangle_k$
\cite{langer} (this should not be confused with the usual dynamical 
structure function, given in terms of the {}Fourier transform of the four-point 
Green function instead). 
We find it more convenient to work with the normalized
structure function, $S_k(t)/S_k(0)$, which, using Eq. (\ref{Gmodes}), is
given by
$
S_k(t)/S_k(0) = N \left(|\chi_1({\bf k},t)|^2 + |\chi_2({\bf
k},t)|^2\right)$.
The wave-number integrated time derivative of the normalized structure 
function  provides the net change 
in the excitations outside the condensate, or the total change of fluctuations
in the condensate. We define this quantity as    
\begin{equation}
\Delta S_{\rm tot} (t) = \left| \int_{k<k_c}  d^3 k \,
\frac{\partial}{\partial t} \left[\frac{S_k(t)}{S_k(0)}\right] \right|\quad .
\label{structure2}
\end{equation}
As shown in Ref. \cite{GG}, the peak of $\Delta S_{\rm tot} (t)$ gives a
measure of the departure from thermal equilibrium following
a quench, while the position of the peak gives the equilibration time
scale of the system. These properties can be extracted from {}Fig. 3,
which shows $\Delta S_{\rm tot} (t)$ for the three
cases defined in {}Fig. 1.

In {}Fig. 3 we observe {\em two peaks} for each temperature
studied. The first peak is clearly related to the time
evolution of the anomalous density (compare with {}Fig. 2), and the second 
peak to the 
time evolution of the condensate density (compare with {}Fig. 1). 
Notice the correlation
of the results regarding the time scale of equilibration for the
condensate (associated to the position of the second peak). 
The lower the temperature of the thermal bath into which the
system is quenched the higher is the peak of $\Delta S_{\rm tot} (t)$.
This is consistent with the fact that the lower the temperature of the 
thermal bath the further from equilibrium the system will be.
The temperature dependence of the equilibration times is quite clear.
{}For low temperatures, $T/T_0 < 0.1$, the first peak
disappears. The equilibration process in this regime is completely
driven by the condensation growth. In this case, the approximations used
in Ref. \cite{BFR} are acceptable since the anomalous density
contribution becomes subleading. On the other hand, for final temperatures
near the critical one, $0.3 < T/T_0 < 1$, 
the second peak in {}Fig.
\ref{fig:structure} disappears and the equilibration process becomes now fully
governed by thermal fluctuations encoded in the mean value of the
anomalous density. Moreover, we find an intermediate temperature region,
$0.1 < T/T_0 < 0.3$ (shown in {}Fig. \ref{fig:structure}), 
where the two equilibration times are present.

In {}Fig. 4 we exhibit the results for the equilibration time
$t_{\rm eq}$ ($\equiv (m a^2 /\hbar) \tau_{\rm eq}$) for both the condensate 
(squares) and the anomalous density (triangles) as a function of the
(dimensionless) density $\rho=n a^3$ at fixed temperature $T/T_0 = 0.1$,
extracted from the location of the second and first peaks, respectively, of 
$\Delta S_{\rm tot}$.
The results can be perfectly fitted by
the function $\tau_{{\rm eq}, \varphi_0^2}^{\rm fit} \simeq 0.2/\rho$ (line),
for the condensate growth, while the result related to the equilibration 
of the anomalous density scales with the density in an
analogous way,  $\tau_{{\rm eq},\langle
\varphi \varphi\rangle}^{\rm fit} \simeq 0.12/\rho$ (dashed line). In 
terms of dimensionful quantities:

\begin{equation}
t_{\rm eq} \simeq 2.5 \frac{\hbar}{ng}\;.
\end{equation}
These results could in principle be
compared with some early qualitative estimates for the growth time of the
condensate for homogeneous gases \cite{kagan}, even though the
physics discussed in those references is essentially different, 
yielding a rough estimate of the
formation time $\sim O[(ng)^{-1}]$. This is exactly the behavior
found here. 

Although we deal with an idealized
situation for the condensation formation, it is useful to 
briefly discuss the experimental results and compare 
them with our estimates,
pointing out clearly the differences and similarities.
Up to date only two experiments have been performed involving
the explicit study of the dynamics of BEC formation.
The first one was performed by the MIT group and reported in 
Ref. \cite{growth1}. They have studied the dynamics of BEC of
atomic $^{23}Na$ ($a=2.75 \; nm$) in a cigar shaped magnetic 
trap, described by frequencies $\omega_x=\omega_y = 2 \pi \, 82.3 \;Hz$
and $\omega_z= 2 \pi \, 18 \; Hz$. The number of atoms in the trap
was $N \simeq 10^7$ and the temperatures involved were between
$0.5 - 1.5\, \mu K$. The geometry of the confining potential yields 
a volume at the center of the trap \cite{reviews}
$V_{\rm trap} = \pi^{3/2} a_x a_y a_z$, where 
$a_i = \sqrt{\hbar/(m \omega_i)}$. The central density can then be
estimated as $n \sim 6.8 \times 10^{22} \; {\rm atoms}/m^3$,
corresponding to $\rho = n a^3 \sim 10^{-3}$. The typical formation
time observed was $\sim 100 - 200 \; ms$.

In the second experiment, Ref. \cite{growth2}, performed
by the Munich group, they used $^{87}Rb$ ($a=5.77 \, nm$)
in a trap characterized by $\omega_x=\omega_y = 2 \pi \, 110 \;Hz$
and $\omega_z= 2 \pi \, 14 \; Hz$. The number of atoms in the trap
was $N \simeq 4.2 \times 10^6$ and the temperatures involved was
$\sim 640 n K$. These parameters lead to a central density value
of $n \sim 2.5 \times 10^{23} \; {\rm atoms}/m^3$,
corresponding to $\rho = n a^3 \sim 5 \times 10^{-2}$. The typical formation
time observed was also $\sim 200 \; ms$.

From the experimental data above, we can estimate in which regime
is each of the experiments. {}For
trapped atom gas experiments $\omega_{\rm exc}$ scales with the trapping
potential frequency and $t_{\rm col}$ is usually a measured quantity. 
Taking $\omega_{\rm exc}\sim \bar{\omega} =
(\omega_x \omega_y \omega_z)^{1/3}$, the average frequency, and using
the experimental data for the relaxation time measured in these
experiments, we find for the MIT experiment that
$\omega_{\rm exc} t_{\rm coll} \sim 3$, while for the Munich experiment
$\omega_{\rm exc} t_{\rm coll} \sim 40$, putting these experiments in
the borderline between the collisionless and hydrodynamical regimes
(MIT) and in the collisionless regime (Munich).
{}For a homogeneous gas it is more difficult 
to get such estimates. However, it is reasonable to estimate
$\omega_{\rm exc}$ as proportional to the thermal energy of the bath $\hbar
\omega_{\rm exc} \sim k_B T$ and $t_{\rm coll}$ expressed in terms of the 
mean time between collisions, which for a classical gas is
$(\sigma \bar{n} \bar{v})^{-1}$ (for an explicitly determination of
the relaxation time, as given by the inverse of the damping rate
for a homogeneous weakly interacting Bose gas, see Ref. \cite{damping}),
where $\sigma = 8 \pi a^2$ is the $s$-wave cross-section, $\bar{n}$
is the maximum non-condensate density and $\bar{v}$ is the characteristic
thermal velocity. Using our parameter values and a temperature of
$T=0.1 T_c$, with $T_c$ approximated by the condensation temperature
for an ideal gas\footnote{{}For an estimate
this is a reasonable approximation 
since for a dilute weakly interacting gas the 
the shift of the transition temperature, $\Delta T_c/T_c$, of
the interacting model compared with the ideal gas transition
temperature, is of order $a n^{1/3}$ \cite{tc}.}, 
we obtain $\omega_{\rm exc} t_{\rm coll} \sim 60$, 
deeply into the collisionless regime, for both $^{23}Na$ and $^{87}Rb$. 
The results for the equilibration times we obtain are
of order $10^{-6}s$ for the parameters of $^{87}Rb$ and
$10^{-7}s$ for  $^{23}Na$.

The large difference in magnitude of time scales 
obtained here as compared to those of the experiments
signals the importance of incoherent/coherent collisional effects 
in the trapped atomic gases experiments due to the nature of the 
experimental setup. Differently from our quenching scenario,
where we can effectively approximate the model by a two-level 
system at low temperatures, in the experiments, 
the confined gas is firstly cooled to a temperature
close to the critical temperature and then it is put 
in a nonequilibrium state
by means of an evaporative cooling through a rapid rf sweep (MIT
experiment) or a constant frequency rf field
(Munich experiment), which removes the highly
excited atoms from the trap. The system in then left to thermalize to a
temperature below $T_c$ and the formation (growth) of the condensate 
occurs. This cannot be approximated by a two-level system. In the
experiments, scattering and growth among adjacent levels, which are
described by those microscopic processes associated with collisions, are
of relevance to the dynamics in this regime. Those processes are indeed
appropriately described through the use of quantum kinetic theory
(quantum Boltzmann equations) for the trapped gases and are necessary to
describe the whole process of evaporative cooling and subsequent
thermalization during the process of condensate formation as shown in
Refs. \cite{stoof,gardiner}.
It is clear that we are considering
a different regime from the beginning. In particular, we do not have a
thermalization regime (since it is considered in thermal equilibrium from
the beginning). Collisional effects lead to a much longer evolution for
the condensate, explaining the disparities of orders of magnitude between
our results and the experimental ones. We recall that, 
as pointed out by Stoof
in Ref. \cite{stoof2}, the regime we studied 
(named the second stage of condensate
formation in that reference) lasts about $O(\hbar/n g)$, 
which corresponds
exactly to the time scale we found, while the relaxation towards
equilibrium in the third and final stage of the condensate formation
(absent in our theoretical setting) leads to a much longer time
evolution for the condensate. 

In summary, we have provided in this paper a way 
to analyze the time scales 
related with the formation of the condensate, 
by studying the behavior of the total
structure function, as defined by Eq. (\ref{structure2}). 
We have analyzed the case
of a homogeneous gas tha undergoes an instantaneous quench and we have 
shown that for temperatures of the 
thermal bath much lower than the critical temperature the gas is deep 
in the collisionless regime, justifying 
the mean field approach developed in this paper. We showed that under 
this circumstances, there are  
two equilibration  time scales perfectly distinguished and associated 
with different physical processes, {\em i.\ e.\ } 
the growth of the condensate and the pair scattering  process  described by the anomalous density.
The structure of equilibration times reported in this paper opens several
questions related to the approach to equilibrium in interacting
dilute gases. For instance, the presence of two equilibration 
time scales may influence time correlations, modifying
fluctuation-dissipation relations. In order to address these issues,
theoretically as well as experimentally, it is necessary to understand
the behavior of correlation functions that depend on two 
different times as, for instance, the
two-time dynamical structure factor. We hope to report on this subject
in the near future. Another issue we are currently pursuing
is the improvement of the equations presented here by including
the higher-order collision terms by means of self-consistent 
Schwinger-Dyson equations for the nonequilibrium 
propagators. These results will be presented in the near future.

\acknowledgments

The authors acknowledge Conselho Nacional de Desenvolvimento
Cient\'{\i}fico e Tecnol\'ogico (CNPq - Brazil) (D.G.B., E.S.F., R.O.R.), 
U.S. DOE -- Contract No. DE-AC02-98CH10886 (E.S.F.), 
``Mr. Tompkins Fund for Cosmology and Field Theory'' at 
Dartmouth (R.O.R.), and NSF -- grants PHY-0070554 and PHY-0099543 (M.G.) 
for financial support. E.S.F. would like to thank the Nuclear 
Theory Group at Brookhaven National Laboratory and the 
Department of Physics $\&$ Astronomy at Dartmouth College, where 
this work was initiated, for their kind hospitality.

%\newpage

\vspace{0.3cm}

\begin{figure}[c]
\centerline{\epsfig{figure=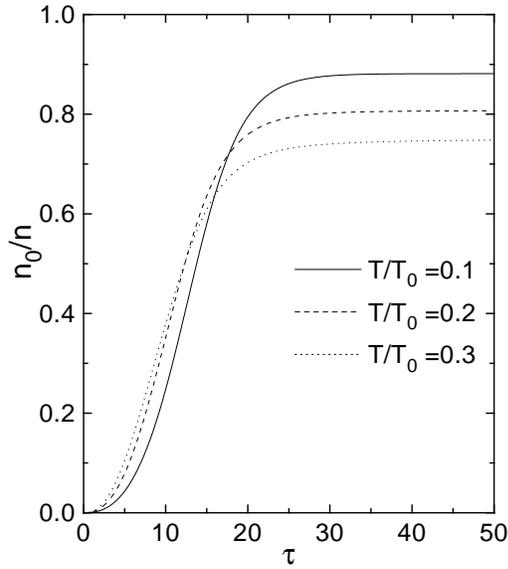,width=6.7cm}}
\caption{Fraction of condensate density as a function of 
time for $na^3=0.01$ and $T/T_0=0.1, 0.2, 0.3$. Here, 
$\tau\equiv (\hbar/ma^2)t$ is a dimensionless time.}
\label{fig:density}
\end{figure}

\vspace{0.5cm}

\begin{figure}[c]
\centerline{\epsfig{figure=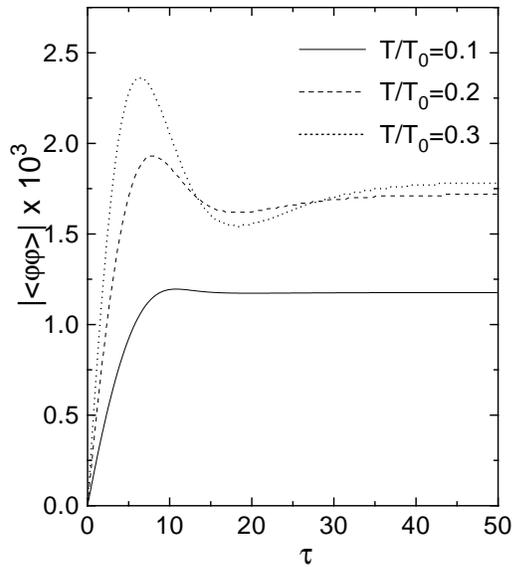,width=6.7cm}}
\caption{Time evolution of the anomalous density 
for the same parameters used in {}Fig. 1.}
\label{fig:anomalous}
\end{figure}

\begin{figure}[t]
\centerline{\epsfig{figure=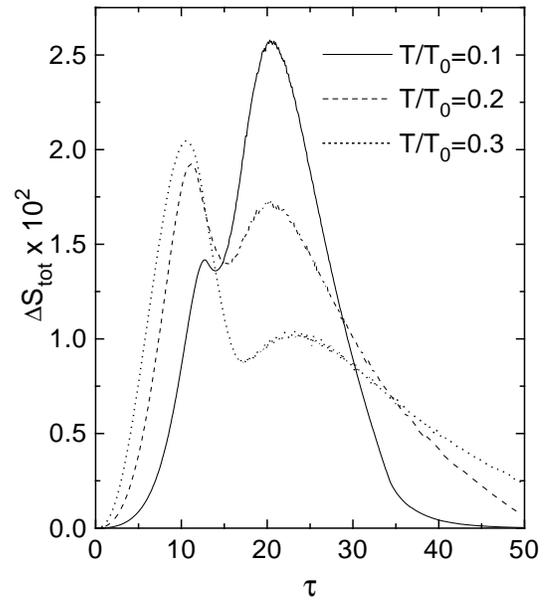,width=7cm}}
\caption{Time evolution of the wave number integrated time derivative
of the normalized structure function for the same parameters used in 
{}Fig. 1.}
\label{fig:structure}
\end{figure}

\begin{figure}[t]
\centerline{\epsfig{figure=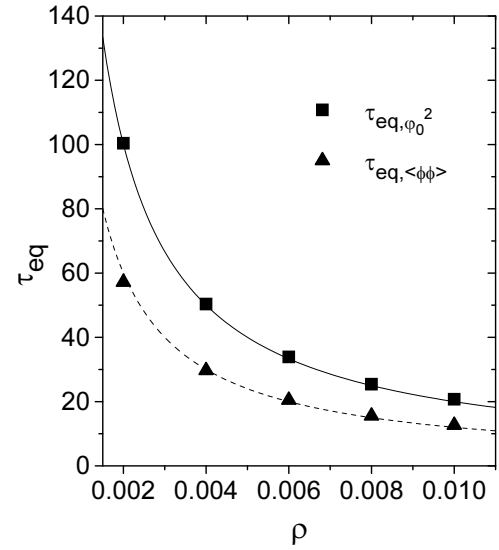,width=7cm}}
\caption{Equilibration time for the condensate (squares) and anomalous
density (triangles) obtained from the second and first 
peaks of  $\Delta S_{\rm tot}$, respectively, as a function
of the density, $\rho= n a^3$, at a fixed temperature $T/T_0=0.1$.}
\label{fig:tau_eq}
\end{figure}

\end{document}